\documentclass[aps,prb,twocolumn,superscriptaddress]{revtex4-1}
\usepackage{graphicx}
\begin{document}

%\preprint{}

\title{Orbital and spin magnetic moments of transforming 1D iron inside metallic and semiconducting carbon nanotubes}

\author{Antonio Briones-Leon}
\email[Correspondence author:]{antonio.briones@univie.ac.at}
\author{Paola Ayala}
\affiliation{Faculty of Physics, University of Vienna, Strudlhofgasse 4, 1090 Vienna, Austria.}

\author{Xianjie Liu}
\affiliation{Department of Physics, Chemistry and Biology (IFM), Link\"oping University, 58333 Link\"oping, Sweden.}

\author{Hiromichi Kataura}
\affiliation{National Institute of Advanced Industrial Science and Technology (AIST), Tsukuba 305-8565, Japan.}

\author{Kazuhiro Yanagi}
\affiliation{Department of Physics, Tokyo Metropolitan University, Hachiouji, Tokyo 192-0397, Japan.}

\author{Eugen Weschke}
\affiliation{Helmholtz-Zentrum Berlin f\"ur Materialien und Energie, Wilhelm-Conrad-R\"ontgen-Campus BESSY II,
Albert-Einstein-Str. 15, 12489 Berlin, Germany.}

\author{Michael Eisterer}
\affiliation{Institute of Atomic and Subatomic Physics, Technische
Universit\"at Wien, Stadionallee 2, 1020 Vienna, Austria}

\author{Thomas Pichler}
\author{Hidetsugu Shiozawa}
\affiliation{Faculty of Physics, University of Vienna, Strudlhofgasse 4, 1090 Vienna, Austria.}

\date{\today}

\begin{abstract}

The orbital and spin magnetic properties of iron inside transforming metallic and semiconducting 1D carbon nanotube hybrids are studied by means of local x-ray magnetic circular dichroism (XMCD) and bulk superconducting quantum interference device (SQUID) measurements. Nanotube hybrids are initially ferrocene filled single-walled carbon nanotubes (SWCNT) of different metallicities. After a high temperature nanochemical reaction ferrocene molecules react with each other to form iron nano clusters. We show that the ferrocene's molecular orbitals interact differently with the SWCNT of different metallicities without significant XMCD response. This XMCD at various temperatures and magnetic fields reveals that the orbital and/or spin magnetic moments of the encapsulated iron are altered drastically as the transformation to 1D Fe nanoclusters takes place. The orbital and spin magnetic moments are both found to be larger in filled semiconducting nanotubes than in the metallic sample. This could mean that the magnetic polarizations of the encapsulated material is dependent on the metallicity of the tubes. From a comparison between the iron 3d magnetic moments and the bulk magnetism measured by SQUID, we conclude that the delocalized magnetisms dictate the magnetic properties of these 1D hybrid nanostructures.

\end{abstract}

%\keywords{XAS, XMCD, magnetic moment, carbon nanotubes, 1D nanostructures}

\maketitle

\section{Introduction}

The extraordinary electronic and mechanical properties of single-walled carbon nanotubes (SWCNT) make them excellent candidates as building components for micro- and nano-devices. The magnetic properties of SWCNT are highly anisotropic due to their unique 1D nanostructure. The diamagnetic nature of carbon nanotubes (CNTs) was predicted theoretically for semiconducting and metallic nanotubes \cite{Ajiki1993JPSoJ}. Later, this was observed experimentally \cite{Lipert2009JoAP}. It was reported that the electronic band structure of SWCNTs is altered at applied magnetic fields parallel or perpendicular to the tube axis due to their anisotropy, leading to novel magnetic, magneto-transport and mangeto-optical properties \cite{Kono2008CNATItSSPaA} elemental for device applications such as high density magnetic recording media. Theoretical and experimental studies have shown that mechanical, chemical and electronic properties can be tuned by different means of functionalization \cite{Hirsch2002ACIE}. In particular, endohedral functionalization or filling of carbon nanostructures with molecules has become promising means to change or even control the electronic and magnetic properties of these hybrid nanostructures \cite{Monthioux2002C}. Since the first observation of peapods, i.e., SWCNT accommodating buckminster fullerenes \cite{Smith1998N}, various molecules and compounds including metallocenes and salts have been encapsulated in the hollow core of CNTs \cite{Sloan2002ICA,Kitaura2007JJoAPP1,Li2005NM,Kitaura2007PRB,Li2006CC,Kitaura2008NR,Li2008JJoAPP1,Shiozawa2008AM,Shiozawa2009PRL,Koshino2010NC,Shiozawa2010AM}. Encapsulated in CNTs, the filling material is protected against oxidation by the rolled up graphene layer. Suggested applications of such materials are magneto recording devices \cite{Geng2006PBM} and nanoscale thermometer for biological purposes \cite{Vyalikh2008N}.

Previous studies on multi-walled carbon nanotubes (MWCNT) encapsulating magnetic nanoparticles (Fe, Ni or Co) grown by chemical vapour deposition (CVD) have shown the magnetic coercivity in contrast to those without catalytic particles inside \cite{Grobert1999APL,Geng2006PBM,Grechnev2010LTP,Ritter2011C,Hisada2011JoMaMM}. A study of magnetic properties of the so-called HiPco nanotubes showed a superparamagnetic behaviour, which was attributed to the remaining catalytic particles \cite{Bittova2011JoPCC}. A ferromagnetic behaviour was observed in Fe@SWCNT even at room temperature, which was explained as a result of a high degree of Fe filling into the nanotubes and the interaction between the Fe nanowires in the bundles of SWCNT \cite{Borowiak-Palen2006CPL}. A theoretical study shows that the local magnetic moment of Fe nanowires encapsulated in SWCNT depends on the size of the Fe nanoparticles, due to the interaction between the particle and the nanotube \cite{Kang2005PRB}. Experimentally, the encapsulation of Fe in SWCNT strongly alter the spin magnetic moment and the magnitude of magnetic anisotropy energy \cite{Wang2008APL}.

Another attempt to alter the magnetic properties of carbon nanotubes, is via the filling of SWCNT with endohedral metallofullerenes \cite{Shinohara2000ROPIP}. Significant changes in magnetic moment of SWCNT encapsulating metallofullerenes (Gd@C$_{82}$ and Dy@C$_{82}$) were found at low temperature (10 K), due to the charge transfer from the metallofullerene to the SWCNT \cite{Kitaura2007PRB,Ayala2011ME}. In addition, SWCNT filled with magnetic salts such as ErCl$_3$ have been studied. The magnetization of purified empty SWCNT were measured and found to be much lower than in ErCl$_3$ nanowires grown into the SWCNT where the magnetization values are the same as the bulk anhydrous ErCl$_3$ \cite{Kitaura2008NR,Ayala2011PRB}.

In recent years, metallocene filled and especially ferrocene filled carbon nanotubes (FeCp$_2$@SWCNT) have been widely studied both experimentally and theoretically. Their electronic properties
can be modified via filling followed by the transformation of the molecules inside into inner tubes and Fe nanoclusters \cite{Shiozawa2008AM, Shiozawa2008PRB} utilizing a nanochemical reaction. By the defined decomposition of the FeCp$_2$ inside the nanotubes, encapsulated Fe nanowire can be formed. Such Fe@SWCNT were reported to show a higher magnetization than the FeCp$_2$@SWCNT and the pristine SWCNT, and exhibit both ferromagnetism and superparamagnetism at different temperatures \cite{Li2008JJoAPP1}.

\par
One of the limitations in the measurement of the magnetic properties in these magnetic 1D nanotructures is the overall "bulk" character, for example by using superconducting quantum interference devices (SQUID). X-ray magnetic circular dichroism (XMCD) spectroscopy allows to unravel  the magnetic states of specified atomic orbitals in compounds and to identify the spin and orbital magnetic moments\cite{Stohr1999JoMaMM}. Hence, XMCD has become a powerful tool for studying the magnetization of a variety of magnetic materials, even in the paramagnetic phase \cite{Shiozawa2003JotPSoJ}. In the case of the study of filled SWCNT hybrids it has been successfully applied to investigate the local magnetic properties of metallofullerene
filled and ErCl$_3$ filled SWCNT \cite{Kitaura2007PRB,Kitaura2008NR}. Yet, none of these studies was done as a function of the metallic character of the SWCNT.

\par
In the present work, we study the local magnetic properties of iron confined in high-purity metallic and semiconducting SWCNT samples as the hosts. The iron was initially introduced inside SWCNT in the form of ferrocene (FeCp$_2$) and transformed to 1D Fe nanoclusters in DWCNT by a thermally driven nanochemical pyrolysis. We show that for the filled starting material the molecular orbital states of ferrocene interact differently with the electronic states of the SWCNT with different metallicities but in both cases we observe no significant XMCD response. In contrast, the nanotubes filled with Fe nanoclusters exhibit enhanced XMCD signals. The orbital ($m_L$) and spin ($m_S$) magnetic moments, evaluated by using the sum rules \cite{Thole1992PRL,Carra1993PRL}, are well fitted by the Langevin function. The $m_L$ of the encapsulated ferrocene is much smaller compared with that of the solid iron. In term, the $m_S$ of the nano iron inside CNT is reduced significantly from the ferrocene value. These results are understood within the framework of the solid state effects and charge transfer which vary depending on the chemical surrounding of the encapsulated iron. Both, the orbital and spin magnetic moments are found to be larger in filled semiconducting nanotubes than in the metallic sample. This could mean that the magnetic polarizations of the encapsulated material are dependent on the metallicity of the SWCNT. Semiconducting and metallic pristine SWCNT have shown a diamagnetic response in SQUID measurements. After the filling, ferromagnetism is observed for the Fe filled metallic SWCNT and paramagnetism for the semiconducting tubes. The coercivity depends on the degree of filling. Much larger positive magnetic moments per iron molecular unit observed by SQUID mean that the delocalized and/or possibly non-iron magnetic polarizations dominate the magnetism of the encapsulated materials, in association with the local moment of the Fe 3d state and the diamagnetism of the SWCNT.

\section{Experimental methods}

\subsection{Samples preparation}

The metallicity-sorted SWCNT samples used in our experiments were synthesized by the arc-discharge process followed by purification, sorting, and film preparation as reported elsewhere \cite{Yanagi2010AN}. The separated SWCNT buckypapers were filled as described in reference \cite{Briones2011PSSB}. The high purity of the SWCNT was confirmed by the X-ray photoemission and C 1s
x-ray absorption (XAS) observations, as well as the 1D characters at the valence-band region, i.e. van Hove Singularities (vHS) and Tomonaga-Luttinger-liquid behaviour, which appear only in SWCNT
with extremely high purity \cite{Ishii2003N,Rauf2004PRL,Ayala2009PRB}. Metallic, semiconducting and mixed nanotube samples were filled with ferrocene. The effective filling was estimated by converting them into double-walled (DW) CNTs by annealing in vacuum at 500$\rm{^o}$C for several hours. Using Raman spectroscopy with a 599 nm excitation wavelength, the radial breathing mode (RBM) of the original SWCNT was observed at the wave number 159 cm$^{-1}$ corresponding to an average diameter of $\sim$1.5 nm. After the annealing treatment, an RBM is observed at around 333 cm$^{-1}$, which corresponds to an inner tube with an average diameter $\sim$0.75 nm of the DW structure, as reported in the literature \cite{Shiozawa2008AM,Sauer2012tbp}. This result proves that the ferrocene is encapsulated in the hollow space of the nanotubes and successively transformed into iron particles by annealing.

\subsection{Measurement of Magnetic Properties}

Fe 2p XAS and XMCD measurements were carried out at the variable polarization undulator beamline UE46-PGM-1 at the BESSY II (Helmholtz-Zentrum Berlin) synchrotron facility. Circular polarization dependent XAS were obtained by measuring the sample drain current with the photon helicity parallel ($\mu^+$) or antiparallel ($\mu^-$) to the sample magnetization. The experimental end-station of this beamline allows cooling the sample down to 5 K at a magnetic field up to 6 T. XAS spectra were taken at the L$_{2,3}$ edge of Fe, with photon energies ranging from the 680 to 750 eV. The base pressure in the measurement chamber was kept below $5\times 10^{-10}$ mbar. A heating station in the preparation chamber was used to in-situ anneal the FeCp$_2$@SWCNT bucky papers at 500$\rm{^o}$C for 10 hrs. Different batches of metallic and semiconducting ferrocene filled SWCNT were annealed in vacuum at 600$\rm{^o}$C for 12 hrs and measured in a SQUID magnetometer MPMS-XL, with magnetic fields up to 7 T and sample temperatures from 5 K to 300 K.

\section{Results and discussion}

\begin{figure}[t]
 \includegraphics[width=\linewidth]{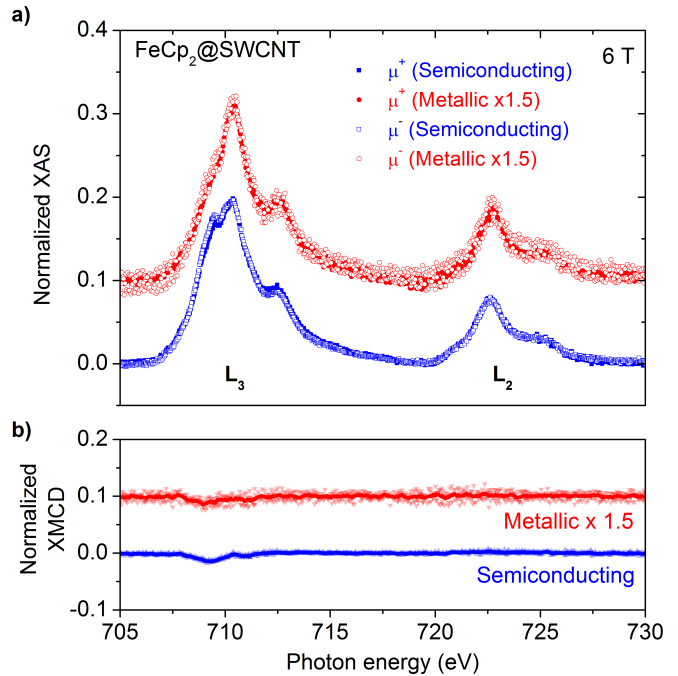}
 \caption{\label{FeCp2atSWCNT} Observed XAS (a) and XMCD (b) response in the L$_{2,3}$ edge of Fe for the metallic and the semiconducting FeCp$_2$@SWCNT samples when a magnetic field of 6 T is applied to the sample at 5 K. All the spectra are offset by 0.1 and normalized to $\int(\mu^++\mu^-)/2$.}
 \end{figure}

The XMCD signal is defined as the difference between the XAS spectra measured at both polarizations ($\mu^+-\mu^-$). We observe no XMCD signals at 0 T (not shown). The XAS and XMCD spectra for the metallicity sorted FeCp$_2$@SWCNT at the L$_{2,3}$ Fe edge at magnetic field of 6 T are depicted in Fig. \ref{FeCp2atSWCNT}. The shown spectra were recorded at 5K. For ease of comparison, the recorded spectra are normalized to the sum of the XAS $\int(\mu^++\mu^-)/2$ and offsetted by 0.1. The XAS response shows slight dependency on the light polarization, Fig. \ref{FeCp2atSWCNT}a. The spectral shapes before annealing are in good agreement with those of ferrocene encapsulated in SWCNT, reported in the previous work on mixed \cite{Shiozawa2008PRB} and separated samples \cite{Sauer2012tbp}. The main spin-orbit splitting features L$_3$ and L$_2$ located at 710.3, 712.5 eV, and 722.6, 724.9 eV, respectively, coincide with those characteristic to the molecular orbitals of ferrocene observed in solids \cite{Hitchcock1990CP}. The feature at 709.5 eV could be characteristic of encapsulated ferrocene, that was less significant but also observed in the previous work \cite{Shiozawa2008PRB}. This feature shows a slight difference between the semiconducting and the metallic samples, which can be attributed to two possible effects: 1) the encapsulated molecules interacting with different proportions of metallic and semiconducting nanotubes or 2) the presence of Fe impurities in the samples. The corresponding XMCD are depicted in Fig. \ref{FeCp2atSWCNT}b. The main peak in the XMCD signal of FeCp$_2$@SWCNT correspond to the low energy band feature at 709.5 eV in the XAS signal. As this feature shows dependency on the metallicity of the SWCNT, the XMCD signal for the semiconducting nanotubes is enhanced from the metallic ones. No contribution from the L$_2$ edge is observed.

Provided that all the spectra were collected at the nearly identical experimental condition, except for the polarization of the light and magnetic field, the spectral intensity integrated over the Fe 2p edge is closely related to the concentration of Fe inside the nanotubes. A high filling factor of 44\% was first determined for a non-separated SWCNT sample from the XAS measurements at the C1s $\pi$ edge and later on with similar amounts for the separated samples by M. Sauer et al. \cite{Sauer2012tbp}. From this value, by considering the differences in the XAS spectral intensity integrated over the Fe edge, filling factors of 49\% and 30\% were determined for the filled semiconducting and the metallic samples, respectively. At room temperature no difference between $\mu^+$ and $\mu^-$ was observed for the semiconducting and metallic FeCp$_2$@SWCNT, which means no XMCD response.

\begin{figure}[t]
 \includegraphics[width=\linewidth]{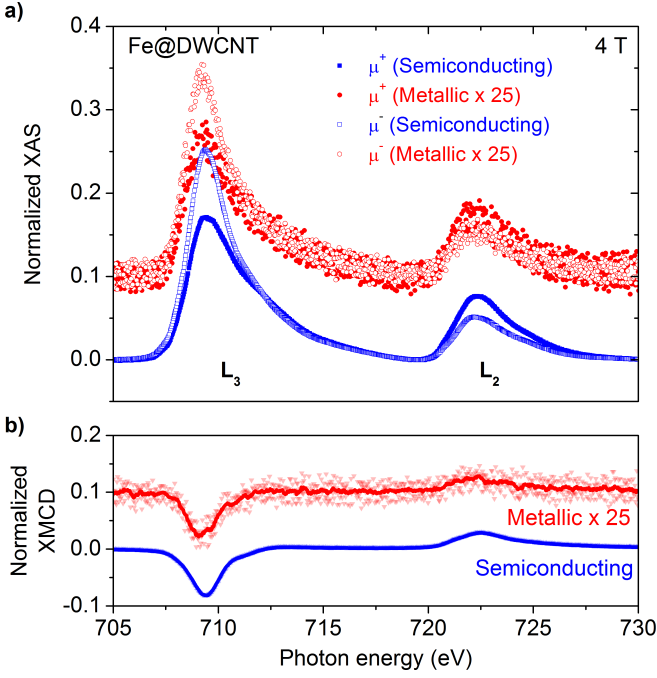}
 \caption{\label{Fe@DWCNT}Observed XAS (a) and XMCD (b) response in the L$_{2,3}$ edge of Fe for the metallic and the semiconducting Fe@DWCNT samples when a magnetic field of 4 T is applied to the sample at room temperature. All the spectra are offset by 0.1 and normalized to $\int(\mu^++\mu^-)/2$.}
 \end{figure}

After annealing the FeCp$_2$@SWCNT samples above 500$\rm{^o}$C for several hours, the ferrocene inside the nanotubes decomposes and forms inner tubes and Fe chains (Fe@DWCNT), as proved by the previous XAS and TEM study \cite{Shiozawa2008AM} and by the inner tubes observed by Raman spectroscopy. In Fig. \ref{Fe@DWCNT}, the XAS and XMCD spectra at 4 T for the samples after the annealing treatment are shown.  The spectra were recorded at room temperature. For ease of comparison, the recorded spectra are normalized to the sum of the XAS $\int(\mu^++\mu^-)/2$ and offset by 0.1. The spectra are significantly altered in shape by annealing and composed of the two main skewed spin-orbit splitting peaks L$_3$ and L$_2$ located at 709.3 and 722.2 eV, respectively, Fig. \ref{Fe@DWCNT}a. From the spectral shapes and energies, these features can be assigned to pure Fe \cite{Chen1995PRL2,Lau2002PRL2}.

The effect of applying a magnetic field is more significant than in the samples before annealing, Fig. \ref{Fe@DWCNT}a. The XMCD signal for the Fe@DWCNT is larger than that for the FeCp$_2$@SWCNT. In Fig. \ref{Fe@DWCNT}b, the formation of the Fe clusters enhances the magnetic response of the filled nanotubes. The difference between the two different metallicity samples is greater for the Fe@DWCNT. This difference can be attributed to 1) the formation of larger Fe clusters due to the high filling degree in the semiconducting nanotubes, and 2) the enhance of the XMCD signal due to the metallicity effect.

\begin{figure}[t]
 \includegraphics[width=\linewidth]{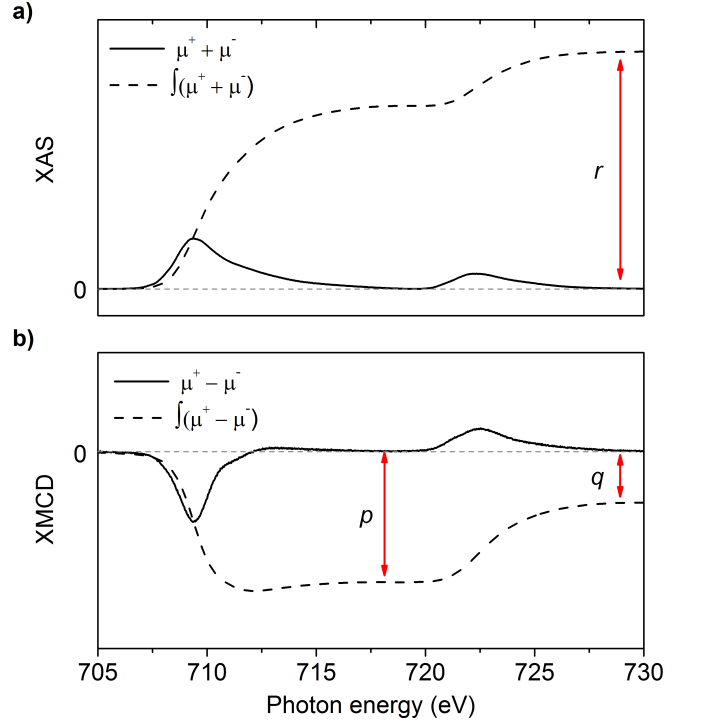}
 \caption{\label{magmom} Definition of the integrals used to calculate the orbital (Eq. \ref{morb}) and spin (Eq. \ref{mspin}) magnetic moments. a) Total XAS and its integral in the whole L$_{2,3}$ edge ($r$); b) XMCD response and its integrals in the L$_3$ ($p$) and the whole L$_{2,3}$ edge ($q$).}
 \end{figure}

The orbital ($m_{orb}$) and spin ($m_{spin}$) magnetic moments have been calculated from the XAS and XMCD spectra by using the sum rules\cite{Thole1992PRL,Carra1993PRL}:

\begin{equation}
m_{orb} =-\frac{4q}{3r}(10-n_{3d})
\label{morb}
\end{equation}
and
\begin{equation}
m_{spin} = -\frac{(6p-4q)}{r}(10-n_{3d})-7\left<T_Z\right>
\label{mspin}
\end{equation}

where $n_{3d}$ is 6.61 corresponding to the 3d electron occupation of Fe calculated theoretically \cite{Guo1994PRB}; $r$ is the integral of the XAS spectrum over the whole L$_{2,3}$ edge, (figure \ref{magmom}a; $p$ and $q$ are the integrals of the XMCD signal at the L$_3$ edge and the whole L$_{2,3}$ edge, respectively (figure \ref{magmom}b). The expectation value of the magnetic dipole operator $\left<T_Z\right>$ in the sum rules is neglected.

\begin{figure}[t]
 \includegraphics[width=\linewidth]{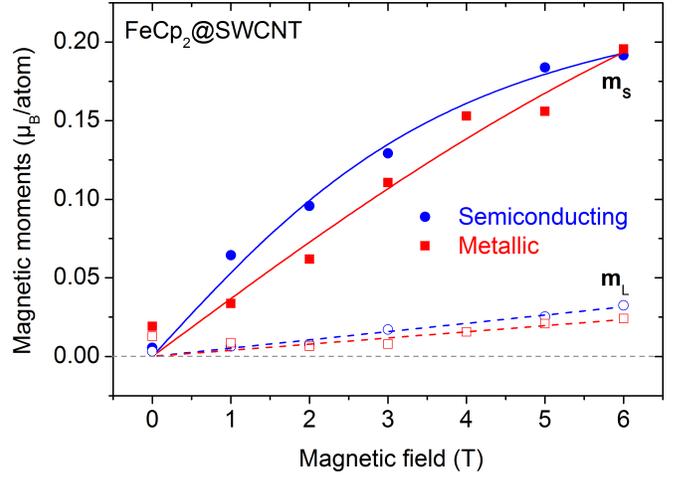}
 \caption{\label{FeCP2magmom}
Spin and orbital magnetic moments of the metallic and semiconducting FeCp$_2$@SWCNT. The data are fitted with the Langevin function. The measurements were done at 5 K.}
 \end{figure}

The experimental spin (m$_S$) and orbital (m$_L$) magnetic moments of the metallicity sorted FeCp$_2$@SWCNT samples are shown in Fig. \ref{FeCP2magmom} as a function of the applied magnetic field. The experimental data were fitted by the Langevin function:

\begin{equation}
M(x) = M_{S,L}(\coth(x)-1/x)
\label{Langevin}
\end{equation}

where $M_{S,L}$ is the saturation magnetization of the spin (S) or the orbital (L) magnetic moments; $x=\mu_cH/k_BT$, in which $\mu_c$ is the uncompensated magnetic moment associated with the nanoparticle core, $H$ is the applied magnetic field, $k_B$ is the Boltzmann constant and $T$ is the temperature.

The obtained saturation values for the spin magnetic moments in the semiconducting ($M_S$=0.26 $\mu_B/atom$) and metallic ($M_S$=0.43 $\mu_B/atom$) FeCp$_2$@SWCNT samples are lower by $\sim$90\% and $\sim$80\%, respectively, than the value reported for pure Fe (2.0 $\mu_B/atom$) \cite{Chen1995PRL2}. The reduced spin magnetic moment in FeCp$_2$@SWCNT could be due to poor interactions between the adjacent Fe atoms, resulting in a low (m$_S$) ordering of the spins. For the orbital magnetic moments, the differences between the bulk Fe and the values obtained for Fe@DWCNT depend on the metallicity of the encapsulating nanotubes. The reported experimental saturation value of the m$_L$ for the pure Fe is $M_L$=0.086 $\mu_B/atom$ \cite{Chen1995PRL2} while we obtain $M_L$=0.0325 and 0.024 $\mu_B/atom$ for semiconducting and metallic nanotubes, respectively. These values correspond to $\sim$38\% and $\sim$28\%, respectively, of the orbital magnetic moment of pure Fe. This reduction in m$_L$ is expected due to the molecular bonding between the Fe atom and the cyclopentadienyl (Cp) rings in the FeCp$_2$ molecule, in accordance with the quenching of the orbital magnetic moment in solids.

By the transformation of the ferrocene to 1D Fe chains inside the nanotubes, the iron 3d magnetic moments increase considerably, Fig. \ref{Fe@DWCNT}. The experimental data was fitted by the Langevin function (Eq. \ref{Langevin}) plus a linear contribution $\chi H$, where $\chi$ is the linear susceptibility. This linear contribution can be attributed to the amorphous character of the Fe chains \cite{Bittova2011JoPCC} and has been reported for other Fe containing molecules \cite{Kilcoyne1995JMMM}.

\begin{figure}[t]
 \includegraphics[width=\linewidth]{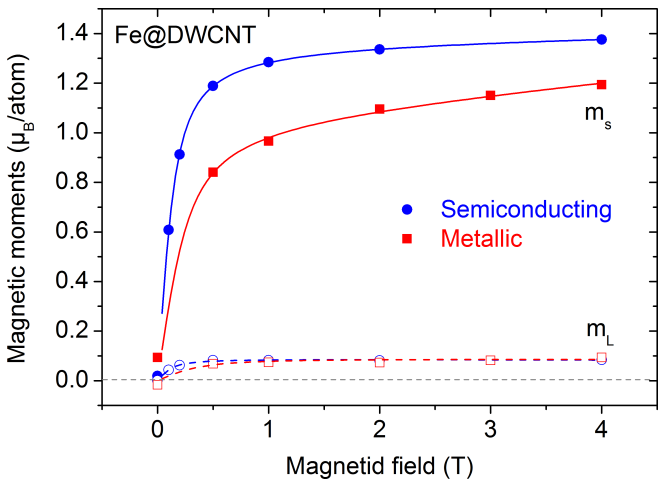}
 \caption{\label{Femagmom}
Spin and orbital magnetic moments of the metallic and semiconducting Fe@DWCNT. The data are fitted with Langevin function. The measurements were done at 300 K.}
 \end{figure}

The saturation value for the orbital magnetic moments of iron in Fe@DWCNT are $M_L$=0.084 and 0.093 $\mu_B/atom$ for semiconducting and metallic, respectively, comparable to those reported for pure Fe (0.086 $\mu_B/atom$)\cite{Chen1995PRL2} with very small differences between the two metallicity samples. The saturation values of the spin magnetic moment for the semiconducting and metallic Fe@DWCNT samples are $M_S$=1.37 and 1.19 $\mu_B/atom$, respectively. These values are lower by $\sim$31\% and $\sim$40\%, respectively, than 2.0 $\mu_B/atom$ reported for pure Fe \cite{Chen1995PRL2}. The saturation values of the spin and orbital magnetic moments for pure Fe and for the filled semiconducting and metallic SWCNT are shown in Table \ref{table_mag_mom}.

The enhanced saturation value after annealing means that the spins are getting ordered as the chemical status of the filling changes from ferrocene to the 1D iron cluster. The difference between the two metalicity Fe@DWCNT samples could be originating from the cluster size due to the filling degree and/or from the metallicity. Since the quantity of available iron atoms most probably defines the mean length of the iron cluster formed inside the nanotube, the higher filling of the semiconducting sample leads to the formation of larger iron clusters which tend to have a greater ordering of spins due to the size effect \cite{Niemeyer2012PRL}. Similarly, the enhanced metallicity of the SWCNT could also reduce the spin ordering in the encapsulated metal via scattering by conduction electrons.

\begin{table}[t]
\caption{Saturation values of the spin (M$_S$) and orbital (M$_L$) magnetic moments in $\mu_B/atom$, of the metallic (M) and semiconducting (SC) FeCp$_2$@SWCNT, Fe@DWCNT and pure Fe.}
\centering
\begin{tabular*}{.48\textwidth}{@{\extracolsep{\fill}} c c c c c c}
%\cline{2-6}
\hline
 & Fe & \multicolumn{2}{c}{FeCp$_2$@SWCNT}  & \multicolumn{2}{c}{Fe@DWCNT} \\
%\cline{3-6}
 &  & SC & M & SC & M \\ 
%\cline{3-6}\hline          
\hline\hline
\multicolumn{1}{c}{M$_S$} & 2.0\cite{Chen1995PRL2} & 0.26 & 0.43 & 1.37 & 1.19   \\ 
\multicolumn{1}{c}{M$_L$} & 0.086\cite{Chen1995PRL2} & 0.0325 & 0.024 & 0.084 & 0.093  \\      % [1ex] adds vertical space
\hline 
\end{tabular*}
\label{table_mag_mom} 
\end{table}

Another batch of metallicity sorted FeCp$_2$@SWCNT and Fe@DWCNT were measured by SQUID in order to compare the "local" and the "bulk" character of the magnetic properties of these materials. The filling inspected by Raman spectroscopy is $\sim$ 34\% and $\sim$8\% for the metallic and semiconducting samples, respectively, in this case.

\begin{figure}[t]
 \includegraphics[width=\linewidth]{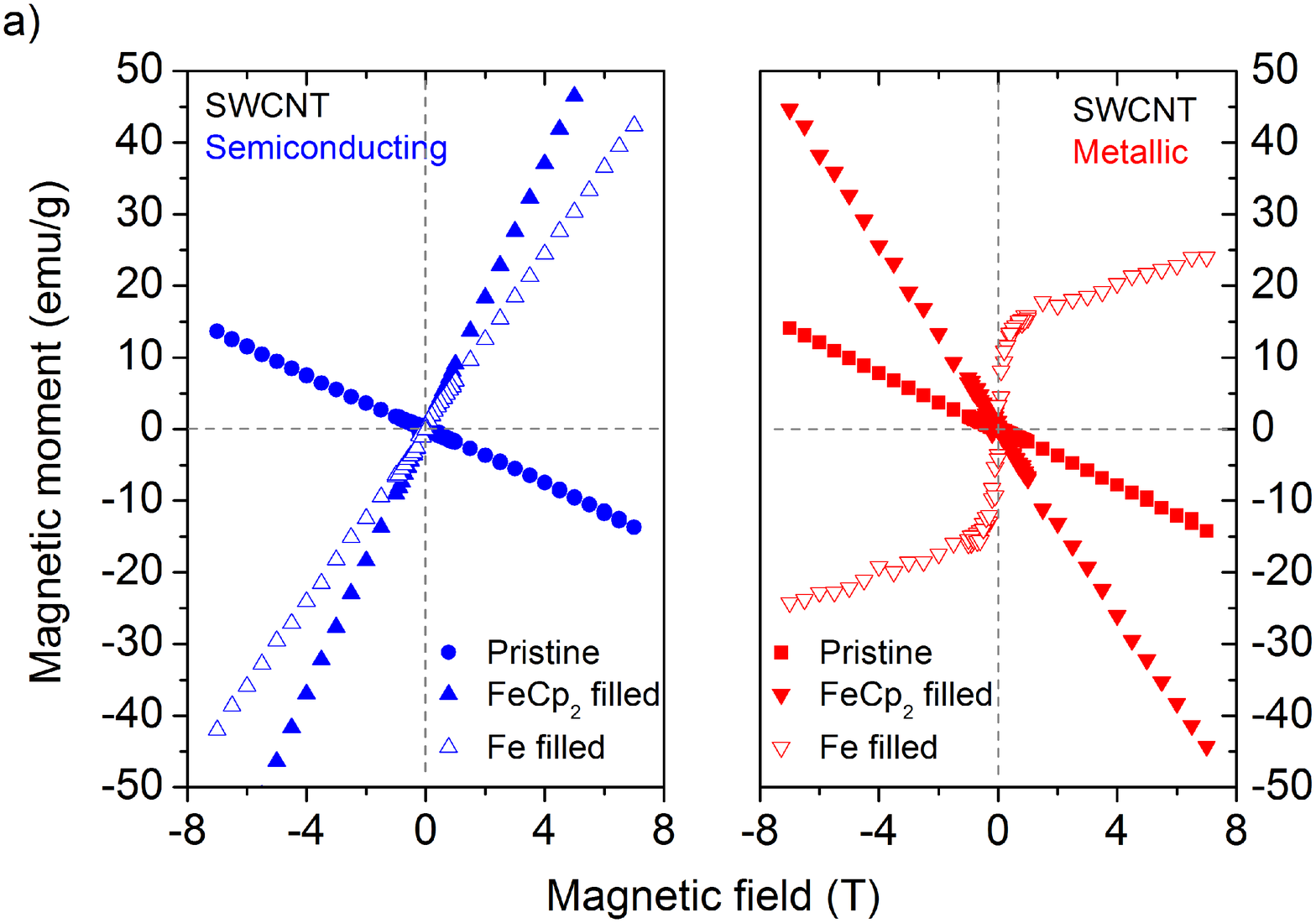}
 \includegraphics[width=\linewidth]{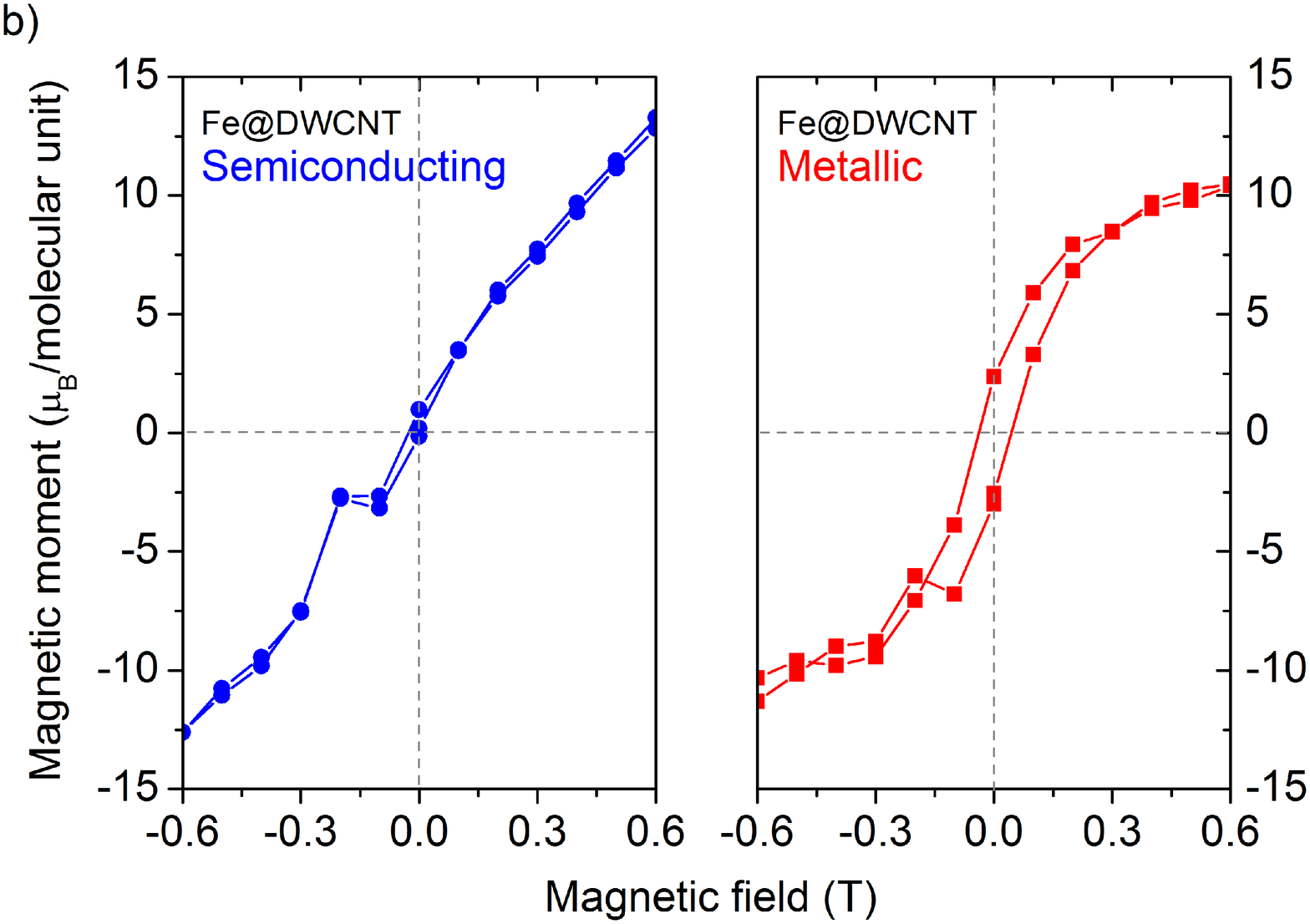}
 \caption{\label{SQUID_fielddep}
Total magnetic moment measured by SQUID. a) For pristine semiconducting and metallic SWCNT, FeCp$_2$@SWCNT and Fe@DWCNT; b) in the low magnetic field region for semiconducting and metallic Fe@DWCNT. The magnetic moments were calculated per molecular unit with one Fe atom (b). The diamagnetic background of the pristine nanotubes were subtracted from the data of the FeCp$_2$ and Fe filled nanotubes. The measurements were done at 300 K.}
 \end{figure}

Figure \ref{SQUID_fielddep} depicts the field dependence magnetization curves for the second batch of semiconducting and metallic samples measured at room temperature. Before the filling, both SWCNT samples show a diamagnetic behaviour, Fig. \ref{SQUID_fielddep}a. This diamagnetism of the pristine SWCNT was subtracted from the FeCp$_2$@SWCNT and Fe@DWCNT.
After the filling of the SWCNT with ferrocene a change from diamagnetism to paramagnetism is observed in the semiconducting sample while the magnetic response of the metallic FeCp$_2$@SWCNT remain diamagnetic, Fig. \ref{SQUID_fielddep}a. The drastic change in the semiconducting sample can be attributed to the following two main aspects: 1) the low interaction between the carbon nanotube and the FeCp$_2$ molecule, and 2) the low filling degree in the semiconducting sample. When both aspects are present, the FeCp$_2$ molecules are electronically isolated and possibly free to rotate inside the tubes. Under a magnetic field, these molecular magnets are aligned so to exhibit the paramagnetic behaviour. In the case of the metallic sample the interaction between the FeCp$_2$ and the SWCNT is strong due to the higher metallicity, as well as between the FeCp$_2$ molecules due to the high filling degree. This can result in the enhanced diamagnetism as observed in the right panel in Fig. \ref{SQUID_fielddep}a.

After the transformation to Fe@DWCNT, the semiconducting sample stays paramagnetic. In the case of the metallic sample, which has a filling degree higher by 23\% than the semiconducting sample, a ferromagnetic behaviour is evident with a coercivity of 40 mT, Fig. \ref{SQUID_fielddep}b. This ferromagnetism obtained is thought to be enhanced due to the larger Fe cluster size and possibly the strong metallic character. The small coercivity value for the semiconducting sample means that the Fe nanoparticles hardly interact between them and the spin alignment is poor owing to the smaller cluster size.

The magnetic moments obtained by SQUID are much larger than those obtained by XMCD, due to the "in bulk" character of the measurements where the contributions of the electron spins in the conduction band are observed. In this case, the remarkable differences between the filled metallic and semiconducting samples observed by SQUID are associated with the delocalized magnetisms which should be more sensitive to changes in SWCNT metallicity.

\section{Conclusion}

The orbital and spin magnetic moments of the iron 3d states in FeCp$_2$@SWCNT and Fe@DWCNT have been studied. A signature for the intermolecular interactions has been observed by XAS on FeCp$_2$@SWCNT. The both orbital and spin magnetic moments are found to be paramagnetic and larger in the filled semiconducting SWCNT than in the filled metallic nanotubes. This can be attributed to differences in iron cluster size that are larger in the semiconducting tubes. Considerable differences between the filled metallic and semiconducting nanotubes have been observed by SQUID. The ferromagnetism observed in metallic Fe@DWCNT and the paramagnetism in filled semiconducting tubes can be explained in accordance with differences in metallicity and cluster size. The delocalized and/or possibly non-iron magnetic polarizations contribute significantly to the magnetic behaviour of these nano composites.

\begin{acknowledgments}

We acknowledge the Helmholtz-Zentrum Berlin - Electron storage ring BESSY II for provision of synchrotron radiation at beamline UE46/PGM-1 and would like to thank Dr. Enrico Schierle for technical assistance. This work was supported by the Austrian Science Funds (FWF), project P621333-N20, and receiving funding from the European Community's Seventh Framework Programme (FP7/2007-2013) under grant agreement n. 226716. P. A. thanks the EU for support from the Discofil project.

\end{acknowledgments}

\end{document}